\documentclass{article}
\usepackage{epsf,epsfig}
\renewcommand{\d}{\partial}
\newcommand{\p}{{\bf p}}
\renewcommand{\k}{{\bf k}}
\newcommand{\bp}{\bar\Phi}
\newcommand{\ph}{\varphi}
\newcommand{\ep}{\varepsilon}
\newcommand{\exv}[1]{\left\langle{#1}\right\rangle}
\renewcommand{\Im}{\textrm{Im}}
\renewcommand{\Re}{\textrm{Re}}
\newcommand{\nn}{\nonumber\\}

\title{Dynamical resummation and damping in the O(N) model}
\author{
Antal Jakov\'ac\footnote{Zolt\'an Magyary fellow,
                             e-mail: jakovac@planck.phy.bme.hu}\\
\scriptsize Dept. of Theoretical Physics,
  Technical University of Budapest, H-1521 Budapest, Hungary}
\date{October 20, 1999}

\begin{document}

\maketitle

\begin{abstract}
  A general real-time formalism is developed to resum the self-energy
  operator of broken symmetry scalar field theories in form of
  self-consistent gap equations for the spectral function. The solution
  of the equations is approximated with finite lifetime quasi-particles.
  In the Landau damping rates viscosity terms, analogous to gauge
  theories, appear, what leads to a finite damping rate for the long
  wavelength Goldstone modes.
\end{abstract}

\section{Introduction}

Scalar theories appear in many physical situation as part of
fundamental theories (eg. Higgs sector, inflaton) or as effective
models (eg. $\sigma$ -- $\pi$ system). Understanding their dynamics is
a precondition to the clarification of various non-equilibrium
phenomena in cosmology and heavy ion physics.

An interesting and probably detectable non-equilibrium process is the
formation and decay of disoriented chiral condensates (DCC) in heavy
ion physics \cite{DCC}. The equations of motion which describe the
system are studied by many authors \cite{DCCstudies} at classical
level and using quantum corrections. The induced current coming from
the polarization of the thermal medium usually is taken into account
at one loop level. Higher loops mean incorporation of quasi-particle
scattering and may have important consequences in certain physical
situations. In gauge theories, for example, a hierarchically organized
two-step one loop analysis resulted in the appearance of viscosity
terms \cite{gaugevisc} in the dynamical characterization of the lowest
energy scales.

\begin{figure}[htbp]
  \begin{center}
    \epsfig{height=3cm,file=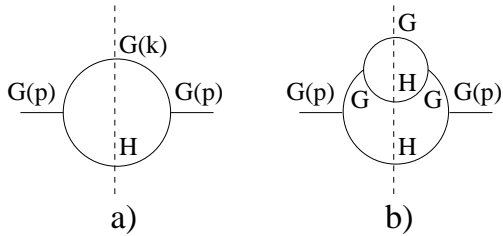}
    \caption{Cut diagrams contributing to Goldstone damping. G means
      Goldstone, H means Higgs modes.}
    \label{fig:goldstdamp}
  \end{center}
\end{figure}

What consequences may we expect from higher loop calculations? Let us
consider the two diagrams relevant for Goldstone damping shown in
Fig.~\ref{fig:goldstdamp}.
The imaginary part of diagram {\em a)} corresponds to the physical
process $G(p)\,G(k)\to H$, where all the particles are on mass shell
and $G(k)$ comes from the heat bath. In this process, if $p\to0$ then
$k$ must go to infinity in order to satisfy the mass-shell relations.
Diagram {\em b)} corresponds to the scattering $G(p)\,H \to G\,H$,
where $H$ comes from the heat bath. Here, even for $p\to 0$ all the
external momenta can be kept finite. In general we will have larger
phase space for the processes contributing to the damping which will
increase the damping rates. This effect can be the most dramatic in
those cases, where the one loop damping rate is small, as in the case
of long wavelength Goldstone modes \cite{smalldamp,ourpaper,BoPaSz}.

The other effect we can expect is due to the decay of the intermediate
particles. At one loop level the intermediate particles are stable,
but higher loops account also for their own damping, what is another
way of interpreting in diagram {\em b)} the inner blob. That means
that there is a finite range in time, where the interactions are
effective. In general this may lead to ``loss-of-memory'' effects for
time scales beyond the typical inverse damping rates (cf. also
\cite{gaugevisc}).

In this paper I would like to take into account the dynamical effects
of a subset of higher loop diagrams by summing up self-consistently
the real time self-energy functions (Section 2). The resulting
generalized gap equations are solved in the quasi-particle
(Breit-Wigner) approximation (Section 3). The results are
applied to the calculation of Goldstone and Higgs Landau damping and,
in particular, the Goldstone on-shell damping rate, at frequencies
much smaller than the width parameters of the resulting Breit-Wigner
shapes (Section 4). The conclusions of the investigation are
summarized in Section 5 where also some important questions left for
future studies are discussed. Some formal tool used for obtaining
results of the main text are presented in an Appendix.

\section{Static and real-time resummations in the spontaneously broken
  phase of the O(N) model}

The Lagrangian of the model of our investigation is written as
\begin{equation}
  {\cal L}(\Phi) = \frac12 (\d\Phi)^2 - \frac12 m^2\Phi^2
  -\frac{g^2}{24} (\Phi^2)^2,
\end{equation}
where $\Phi$ represents a vector in the N-dimensional space
$\Phi=(\Phi_1,\dots,\Phi_N)$. In the symmetry broken phase, where
$m^2<0$, the field has a non-vanishing expectation value $\exv{\Phi} =
\bp=$const, which can be chosen to point in the $\Phi_1$ direction.
We introduce a new field $\ph=\Phi-\bp$, and the Lagrangian takes the
form
\begin{equation}
  {\cal L} ={\cal L}(\bp) - {\cal L}_{tadpole} + \frac12 \ph K^0(i\d,m)
  \ph -\frac{g^2}6\bp\ph_1\ph^2 -\frac{g^2}{24}(\ph^2)^2,
\end{equation}
where the tadpole terms are
\begin{equation}
  {\cal L}_{tadpole}= m^2 \bp\ph_1 + \frac{g^2}6\bp^3 \ph_1,
\end{equation}
and the kinetic term is determined by the kernel
\begin{equation}
  K^0_{ab}(i\d,m)= \left((i\d)^2 -m^2 -\frac{g^2}6\bp^2\right)
  \delta_{ab} -\frac{g^2}3 \bp^2 \delta_{a1}\delta_{b1}.
\end{equation}
We will omit from the Lagrangian the purely background dependent and
tadpole terms in the sequel.

It is well known that the negative mass squared leads to IR
instabilities, which appear also in static quantities, for example as
a complex free energy density for small $\bp$'s. In order to avoid
these effects one has to resum the perturbation series, which allows
for the use of the physical mass in the propagators \cite{resum}. A
consistent method of static resummation consists of introducing
thermal counterterms
\begin{equation}
   {\cal L} =\frac12 \ph K^0(i\d,m_T) \ph -\frac{g^2}6 (\bp\ph)\ph^2
   -\frac{g^2}{24}(\ph^2)^2 + \frac12 (m_T^2-m^2) \ph^2,
\end{equation}
where the last term is also treated as interaction. The determination
of the thermal mass can be done with help of gap equations: the
masses (for example $G^{-1}(k=0)$) are stable against loop corrections
\cite{gapeq}. Since the thermal counterterm preserves $O(N)$ symmetry
the Goldstone theorem ensures the masslessness of the Goldstone bosons
in the broken symmetry case (also at finite temperature
\cite{ourpaper,Goldth}). On the other hand the Higgs mass calculated
from the tree level and one loop formulae should coincide. These
requirements lead to
\begin{eqnarray}
  && 0 = m^2 + \frac{g^2}6\bp^2 + \Pi_i(k=0,m=m_T),\nn
  && m_T^2 = m^2 + \Pi_1(k=0,m=m_T).
\label{massresum}
\end{eqnarray}
These are two equations for the two quantities $\bp$ and $m_H$. In
more complicated cases (especially in gauge theories) we can rearrange
the perturbation theory in a momentum dependent way \cite{momdepres}.

In dynamical systems there are more scales than in the static case. In
the $O(N)$ model besides the temperature ($T$) and the Debye mass
($gT$) there appear the on-shell damping rates for the Higgs and
Goldstone bosons ($g^2T$). Especially in the Goldstone case, where the
Debye scale is missing ($m_G=0$) can the damping play an important
role.  Also at low external momenta (both $k_0$ and $k$ falling for
instance into the range of Landau damping) these scales may have an
impact on the results. Therefore it is worth to work out a consistent
method for the incorporation of the damping scales into the
resummation.

On-shell damping arises when the mass shell falls on some cut of the
propagator. The cuts give momentum dependent contribution to the
self-energy, which suggest to work with a momentum dependent
resummation scheme
\begin{equation}
  {\cal L} =\frac12 \ph K(i\d) \ph -\frac{g^2}6 (\bp\ph)\ph^2
  -\frac{g^2}{24}(\ph^2)^2 + \frac12 \ph P(i\d) \ph,
\end{equation}
where $K(i\d)= K^0(i\d) - P(i\d)$. The counterterm has to preserve
$O(N)$ symmetry at least for the zero momentum modes in order to be
able to satisfy the requirements $\exv{\ph}=0$ and $m_G^2=0$
simultaneously to all orders of the resummed perturbation theory. It
also has to maintain $O(N-1)$ symmetry of the Goldstone modes for all
momenta.

At finite temperature in the real time formalism one works with
several propagators ($G^<,\, G^>,\, G^c, G^a$), and we have to relate
them to the kinetic term in the Lagrangian. The most comfortable way
to do this is first to express the spectral function
$\rho_{ij}=\exv{[\ph_i(x),\ph_j(0)]}$ in momentum space. It is
diagonal $\rho_{ij}(p)=\delta_{ij} \rho_i(p)$, and
\begin{equation}
  \rho_{i}(p) = \frac{-2\Im K_i(p)}{(\Re K_i(p))^2 + (\Im K_i(p))^2)},
\end{equation}
where $\Im K(p)$ is the shorthand notation for $\lim_{\ep\to 0} \Im
K(p+i\ep)$, and similarly for $\Re K(p)$. $\Im K(p)$ has nonzero
values along the cut of the propagator. The $O(N-1)$ symmetry of the
Goldstone bosons ensures the $i$-independence of $\rho_{i\neq1}$.  We
can express the propagators through the spectral function as follows
\cite{LeBellac}
\begin{equation}
  \begin{array}[c]{ll}
    iG^<_i(p) = n(p_0)\rho_i(p) & iG^>_i(p) = (1+n(p_0)) \rho_i(p) \cr
    iG^c_i(t,\p) = \Theta(t) \rho_i(t,\p) + iG^<_i(t,\p)\quad &
    iG^a_i(t,\p) = iG^>_i(t,\p) - \Theta(t) \rho_i(t,\p).\cr
  \end{array}
\end{equation}

\begin{figure}[htbp]
  \begin{center}
    \epsfig{height=2.5cm,file=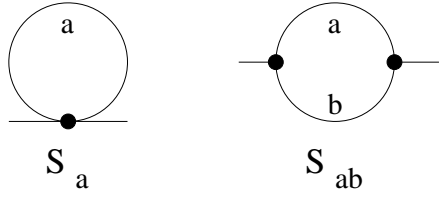}
    \caption{Self-energy contributions}
    \label{fig:pir}
  \end{center}
\end{figure}

Now we are in the position to develop perturbation theory with
``resummed'' propagators. The Feynman rules are unchanged (cf.
\cite{LeBellac}), only the propagators are modified. We are interested
in the resummed, retarded self-energies ($i=2,\dots,N$ and
$a,b=1,\dots, N$) for which the one-loop contributions can be written
with help of two types of Feynman integrals (cf. Appendix)
\begin{eqnarray}
  \Pi^R_1(k) && = \frac{g^2}2 S_1 + \frac{g^2}6 (N-1) S_i +
  \frac{g^4}2 \bp^2 S_{11}(k)+ \frac{g^4}{18} \bp^2 (N-1)
  S_{ii}(k)\nn
  \Pi^R_i(k) && = \frac{g^2}6 S_1 + \frac{g^2}6 (N+1) S_i +
  \frac{g^4}9 \bp^2 S_{1i}(k).
\label{pirs}
\end{eqnarray}
The physical masses are
\begin{equation}
  m_H^2= m^2+\frac {g^2}2\bp^2 + \Pi_1(0), \qquad 
  m_G^2= m^2+\frac {g^2}6\bp^2 + \Pi_i(0).
\end{equation}
The quantities $S_a$ and $S_{ab}$ correspond to the diagrams shown
in Fig.~\ref{fig:pir}.
$S_a$ is momentum independent, it yields merely a modification to the
mass. This contribution is proportional to $G^c(x=0) = G^<(x=0)$,
since $\rho(x=0)=0$. In Fourier space it reads
\begin{equation}
  S_a := \int\frac{d^4p}{(2\pi)^4}\,n(p_0)\rho_a(p).
\end{equation}
The other diagram is nonlocal, especially it has
imaginary part in the momentum representation. In the coordinate space
\begin{eqnarray}
  S_{ab}(x) && = (iG^c_a(x))(iG^c_b(x)) - (iG^<_a(x))(iG^<_b(x)) \nn
  && = \Theta(t) [\rho_a(x)\rho_b(x) + \rho_a(x) iG^<_b(x)+ \rho_b(x)
  iG^<_a(x)].
\end{eqnarray}
In Fourier space it has the form
\begin{equation}
  S_{ab}(k) = \int\!\frac{dp_0}{2\pi} \frac{dp'_0}{2\pi}
  \frac{d^3\p}{(2\pi)^3}\,\frac{\rho_a(p_0,\p) \rho_b(p'_0,\k-\p)}
  {k_0-p_0-p'_0+i\ep}(1+n_a(p_0)+ n_b(p'_0)).
\label{sabdef}
\end{equation}
Its imaginary part with respect to the $k_0$ variable is given by
\begin{equation}
  \Im S_{ab}(k) = -\frac12 \int\!\!\frac{d^4p}{(2\pi)^4}
  \frac{d^4p'}{(2\pi)^4} \,\rho_a(p) \rho_b(p') (1+n_a(p_0)+
  n_b(p'_0)) (2\pi)^4 \delta(k-p-p'). 
\end{equation}
Using
\[ 1+n_a(p_0)+ n_b(p_0) =(1+n_a(p_0))(1+ n_b(p'_0)) -
n_a(p_0)n_b(p'_0) \] 
the above result can be interpreted as the difference of pair creation
rates into (and annihilation out of) the states $a$ and $b$ where the
density of states are $\rho_a$ and $\rho_b$ \cite{Mahan}.

The expectation value for the field operator can be worked out very
simply. $\exv{\ph_i}=0$ is trivially fulfilled, $\exv{\ph_1}=0$
requires
\begin{equation}
  \bp \left(m^2 +\frac{g^2}6 \bp^2 + \frac{g^2}2 S_1 +
    \frac{g^2}6 (N-1) S_i\right) =0.
\label{ph1exv}
\end{equation}
Using this relation with $m_G^2= 0$ leads to
\begin{equation}
  -S_1 + S_i +\frac{g^2}3 \bp^2 S_{1i}(k=0) =0.
\end{equation}
This is fulfilled in the perturbation theory \cite{ourpaper}, but in a
general resummation might fail. In order to maintain the Ward
identities we have to make some further efforts, and perform also
vertex resummation \cite{vertres}. In this context it is sufficient to
add and subtract from the original Lagrangian a piece of the form
$\Delta g^2 \Phi_1^4/24$, where $\Delta g^2\sim g^4$. The original
Lagrangian does not change, but, if we take into account the two
pieces at different loop orders of the perturbation theory, their
contributions break the $O(N)$ symmetry at each level. At tree level
the Goldstone mass is not affected, but in the second term of
(\ref{ph1exv}) leads to the substitution $g^2\bp^2 \to (g^2+\Delta
g^2)\bp^2$. At one loop the compatibility of $\exv{\ph}=0$ and $m_G=0$
is fulfilled if
\begin{equation}
  -\frac{\Delta g^2}{2g^2}\bp^2 -S_1 + S_i +\frac{g^2}3 \bp^2
  S_{1i}(k=0) =0,
\end{equation}
which determines $\Delta g^2$. Keeping this in mind we omit vertex
resummation, since it has no effect up to the order of the
perturbation theory which is considered.

Introducing the notations
\begin{eqnarray}
  && m_T^2 = m^2 +\frac{g^2}2 S_1 + \frac{g^2}6 (N-1) S_i,\nn
  && g_T^2 = g^2 \left( 1 + \frac{g^2}2 S_{11}(k=0) +
  \frac{g^2}{18}(N-1) S_{ii}(k=0)\right)
\end{eqnarray}
($g_T^2$, as can be easily checked, is the one-loop corrected coupling)
we have the relations
\begin{equation}
  \bp^2 = \frac{-6m_T^2}{g^2},\qquad m_H^2 = \frac{g_T^2}3 \bp^2 =
  -2\frac{g_T^2}{g^2} m_T^2.
\label{bpmh}
\end{equation}
Therefore to the given order we can substitute $g^4\bp^2$ by $3 g^2
m_H^2$.

For the optimization of the time-dependent perturbation theory we can
use the same ideas as in the static thermal counterterm case: we
demand that the one loop propagator be identical with the tree level
one for all momenta. Implementing this requirement in the
Schwinger-Dyson form for the inverse propagator
\begin{equation}
  G^{-1}(p) = K(p) - \Pi(p;K) +P(p) = K(p)
\end{equation}
yields
\begin{equation}
  P(p)=\Pi(p;K),
\end{equation}
ie. the dynamical counterterm is the self-energy calculated using the
one-loop exact propagator. Earlier we have determined the expressions
for the retarded self-energies as functionals of the spectral
functions, and also the spectral function is expressed through the
kinetic term $K$.  Thus we can write down a complete set of
renormalized functional equations
\begin{eqnarray}
  \Pi_1(k) && = \frac{g^2}2 S_1 + \frac{g^2}6 (N-1) S_i +
  \frac{3g^2}2 m_H^2 S_{11}(k) + \frac{g^2}{6} m_H^2 (N-1)
  S_{ii}(k)\nn
  \Pi_i(k) && = \frac{g^2}6 S_1 + \frac{g^2}6 (N+1) S_i +
  \frac{g^2}3 m_H^2 S_{1i}(k)\nn
  \rho_1(k) && = \frac{-2 \Im \Pi_1(k)}{(k^2-m_H^2 - \Re (\Pi_1(k) -
  \Pi_1(0)))^2 + (\Im \Pi_1(k))^2}\nn
  \rho_i(k) && = \frac{-2 \Im \Pi_i(k)}{(k^2 - \Re(\Pi_i(k) -
  \Pi_i(0)))^2 + (\Im \Pi_i(k))^2},
\label{gapeq}
\end{eqnarray}
where $m_H$ comes from eq.~(\ref{bpmh}). This set of functional
equations can be, in principle, solved. Their construction represents
the main conceptual result of our investigation. Now, we shall proceed
with developing practical approximation schemes applicable to these
equations.

\section{The Breit-Wigner approximation}
\label{sec:eval}

We can simplify the solution of our gap equations in (\ref{gapeq})
with some assumptions. The most severe is the quasi-particle
approximation, that is we assume that numerically the most important
contribution comes from a narrow region around the mass shell
$\omega(\k)$ of the spectral function (cf. \cite{BWapprox}). This
assumption means that we take into account only those effects which
stem from the resonant broadening of the mass-shell. If contributions
coming from the momentum ranges of the spectral function which are far
from the single particle mass shell prove to be important, then our
Breit-Wigner approximation is not valid any more. A useful
characteristics of the actual situation is the value of the wave
function renormalization constant $Z_\k$ \cite{Zmean}. If its value is
far from 1, the quasi-particle approximation fails, while for
$Z\approx 1$ we can hope that we are on the right track. Formally this
assumption corresponds to the Breit-Wigner approximation to $\Pi(k)$,
with $\Im\Pi(k)\approx -2\omega_\k \gamma_\k$ independent on $k_0$.
For $\gamma\ll \omega$ we can write
\begin{equation}
  \rho_a^{BW}(p)=\frac\pi{\omega_a(\p)}\left(\delta_{\gamma_a(\p)}
  (p_0-\omega_a(\p)) - \delta_{\gamma_a(\p)}(p_0+\omega_a(\p))\right),
\label{BWrho}
\end{equation}
where $\omega_a(\p)$ will be approximated by the mass-shell relation as
\begin{equation}
  \omega_i^2(\p)\approx \p^2 + m_a^2,
\end{equation}
$m_a^2$ standing for the resummed masses (cf. (\ref{massresum})), ie.
we neglect the momentum dependent real part of the self-energies.
Furthermore,
\begin{equation}
  \delta_\gamma(x) = \frac1\pi \Im\frac 1{x-i\gamma} = \frac1\pi
  \frac\gamma{x^2+\gamma^2}
\end{equation}
is a smeared out delta function, where the on-shell damping rates
$\gamma_i(\p)$ yield the width of the functions. In the free case we
get back to the usual expression
\begin{equation}
  \lim\limits_{\gamma\to 0^+} \rho_i^{BW}(p)= (2\pi)\ep(p_0)
  \delta(p^2-m_i^2).
\end{equation}
Under these approximations we can simplify the gap equations using
instead of the last two equations of (\ref{gapeq}) eq.~(\ref{BWrho})
with the usual form of the on-shell damping
\begin{equation}
  \gamma_1(\p) = \frac{-\Im \Pi_1^R(\omega_1(\p),\p)}{2\omega_1(\p)}
  ,\qquad   \gamma_i(\p) = \frac{-\Im \Pi_i^R(|\p|,\p)}{2|\p|}.
\label{gammagap}
\end{equation}

As a further approximation we treat $\delta_\gamma$ acting on any
function smooth around the mass shell as an ordinary Dirac-delta
distribution. This is appropriate if the rate of variation of the
smooth function is slower than (or at least comparable with)
$\gamma$. Since $\gamma$ is the smallest scale in the system
($\gamma\sim g^2 T$) we can apply it in many cases. If we substitute
the Bose distribution with its on-shell value, $S_a$ turns out to be
$\gamma$-independent
\begin{equation}
  S_a\approx \int\frac{d^3\p}{(2\pi)^3}\,\frac1{2\omega_a}\,[1+2
  n(\omega_a)]. 
\end{equation}
For the computation of $S_{ab}$ we introduce the notations
\begin{eqnarray}
  && \p'=\k-\p,\quad\gamma=\gamma_a(\p),\quad \gamma'=\gamma_b(\p'),
  \quad \omega=\omega_a(\p),\quad\omega'=\omega_b(\p')\nn
  && n=n(\omega),\quad n'=n(\omega'),\quad\delta_a^\pm = \delta_a
  (p_0\pm\omega)\quad \delta_b^\pm = \delta_b(p'_0\pm\omega'),
\end{eqnarray}
which allows us to write (cf. (\ref{sabdef}) and (\ref{BWrho}))
\begin{eqnarray}
  S_{ab}(k)=\int\!\frac{dp_0}{2\pi} \frac{dp'_0}{2\pi}
  \frac{d^3\p}{(2\pi)^3}\, \frac{\pi^2}{\omega\omega'}\,\biggl[ &&
  \frac{\delta_a^-\delta_b^- - \delta_a^+ \delta_b^+}
  {k_0-p_0-p'_0+i\ep}\,(1+n+n') \nn&&+ \frac{\delta_a^-\delta_b^+ -
  \delta_a^+ \delta_b^-}{k_0-p_0-p'_0+i\ep}\,(n'-n)\,\biggr].
\end{eqnarray}
Now we can perform the $p_0$ and $p'_0$ integrals with help of the
formula (cf. Appendix)
\begin{equation}
  \frac1{2\pi}\int\limits_{-\infty}^\infty\frac{dx}{(x+i\alpha)
    (\Omega -x +i\gamma)} =  \ep(\gamma) \Theta(\gamma
    \alpha)\,\frac{-i}{\Omega+ i(\alpha+\gamma)}. 
\label{intformula}
\end{equation}
This yields
\begin{eqnarray}
  \int\limits_{-\infty}^\infty\!dp_0\,\frac{ \delta_a(p_0\pm\omega)}
  {\Omega-p_0+i\ep} && = \int\limits_{-\infty}^\infty\!dp_0\,\frac1{2\pi
  i}\left( \frac1{p_0\pm\omega-i\gamma} - \frac1{p_0\pm\omega+
  i\gamma} \right) \frac1{\Omega-p_0+i\ep} \nn&&=
  \frac12\,\frac1{\Omega \pm \omega +i(\ep+\gamma)},
\end{eqnarray}
ie. $\delta_a$ behaves almost like an ordinary Dirac-delta in this
case, but its width is added to the imaginary part of the
denominator. The same formula can be applied again to the $p_0'$
integration, and finally we arrive at
\begin{eqnarray}
  S_{ab}(k)= \int\!\frac{d^3\p}{(2\pi)^3}\,\frac1{4\omega\omega'}
  \biggl[&& \frac{1+n+n'}{k_0-\omega-\omega'+i\Gamma} -
  \frac{1+n+n'}{k_0+\omega+\omega'+i\Gamma} \nn&&+
  \frac{n-n'}{k_0+\omega-\omega'+i\Gamma} - 
  \frac{n-n'}{k_0-\omega+\omega'+i\Gamma}  \biggr],
\label{SGam}
\end{eqnarray}
where $\Gamma=\gamma+\gamma'$. 

The result formally corresponds to changing the imaginary piece $i\ep$
in the perturbative denominators to $i\Gamma$, which has the physical
content of replacing the adiabatic switching off the interactions by
the effect of the particle decay. This leads to the realization of our
conjecture in the Introduction, namely the loss-of-memory effects.
After a time $\sim 1/\Gamma$ the interactions become ineffective and
we can expect in the time evolution of any physical quantity the
appearance of effectively localized (instead of long tail) kernels.

\section{Imaginary part of the self-energies}

The other expected effect, the larger available phase space of final
states contributing to quasi-particle damping will be demonstrated via
the actual calculation of the imaginary parts. From eq.~(\ref{gapeq})
we conclude 
\begin{eqnarray}
  && \Im \Pi_1^R(k) = \frac{3g^2}2 m_H^2 \Im S_{11}(k)+ \frac{g^2}6
  m_H^2 (N-1) \Im S_{ii}(k)\nn 
  && \Im \Pi^R_i(k) = \frac{g^2}3 m_H^2 \Im S_{1i}(k).
\label{impirs}
\end{eqnarray}
Then eq.~(\ref{SGam}) gives
\begin{eqnarray}
  \Im S_{ab}(k) = -\int\!\frac{d^3\p}{(2\pi)^3}\,\frac\pi{4\omega\omega'}
  \biggl[&& (1+n+n')(\delta_\Gamma(k_0-\omega-\omega') -\delta_\Gamma
  (k_0+\omega+\omega')) \nn&& +\, (n-n')(\delta_\Gamma(k_0+\omega-
  \omega') - \delta_\Gamma(k_0-\omega+\omega')) \biggr].\nn
\end{eqnarray}
We can use again our assumption that $\delta_\Gamma$, acting on a
smooth function behaves as an ordinary Dirac-delta function. Here we
apply it to Bose-Einstein distributions as well as to damping rates.
The latter are functions of $|p|$, or $\omega_a$, thus we replace the
argument of $\gamma(\omega')$ by the value dictated by the Dirac-delta
condition. We use also the identity $1+n(x)+n(-x)=0$, and we obtain
\begin{eqnarray}
  \Im S_{ab}(k) = -\int\!\frac{d^3\p}{(2\pi)^3}\,\frac\pi{4\omega\omega'}
  \bigl[&&
    (1+n(\omega)+n(k_0-\omega))\delta_\Gamma(k_0-\omega-\omega') \nn&& 
  -\,(n(\omega) -n(k_0+\omega))\delta_\Gamma(k_0+\omega+\omega') \nn&& 
  +\,(n(\omega) -n(k_0+\omega))\delta_\Gamma(k_0+\omega-\omega') \nn&& 
  +\,(n(\omega-k_0) -n(\omega))\delta_\Gamma(k_0-\omega+\omega') \bigr].  
\end{eqnarray}
The integrand depends only on $\omega$ and $\omega'$, therefore it is
worth to change the integration measure correspondingly. In spherical
polar coordinates $d^3\p = 2\pi dp\,p^2 dx$, where $x=\hat\p\hat\k$.
We change variables from $p$ to $\omega$ and from $x$ to $\omega'$.
Then we obtain in a straightforward procedure
\begin{equation}
  \int\!\frac{d^3\p}{(2\pi)^3} \,\frac\pi{4\omega\omega'}\,
  f(\omega,\omega') = \frac1{16\pi|\k|}\int\limits_{m_a}^\infty
  d\omega \int\limits_{\omega_-}^{\omega_+} d\omega'\,
  f(\omega,\omega'),
\end{equation}
where
\begin{equation}
  \omega_\pm^2 = (p\pm|\k|)^2 + m_b^2,\qquad \omega^2=p^2+m_a^2.
\end{equation}
The sole $\omega'$-dependence remained in the $\delta_\Gamma$
functions. Introducing smeared theta functions via the relations
\begin{equation}
  \int\limits_{-\infty}^x\,dy\,\delta_\Gamma(y) = \Theta_\Gamma(x) =
  \frac12 + \frac1\pi \arctan\frac x\Gamma
\label{smtheta}
\end{equation}
($\lim_{\Gamma\to0}\Theta_\Gamma(x)=\Theta(x)$) and
\begin{equation}
  \Theta_\Gamma(a<x<b) = \Theta_\Gamma(x-a) - \Theta_\Gamma(x-b) =
  \frac1\pi \left(\arctan\frac{(x-a)}\Gamma -\arctan\frac{(x-b)}
  \Gamma \right)
\label{comptheta}
\end{equation}
we immediately find
\begin{eqnarray}
  \Im S_{ab}(k) = -\frac1{16\pi|\k|} \int\limits_{m_a}^\infty\!d\omega
  \bigl[&&
  (1+n(\omega)+n(k_0-\omega))\Theta_\Gamma(\omega_-<k_0-\omega
  <\omega_+)  \nn&& -\,(n(\omega) -n(k_0+\omega))\Theta_\Gamma(\omega_-<
  -k_0-\omega<\omega_+) \nn&& 
  +\,(n(\omega) -n(k_0+\omega))\Theta_\Gamma(\omega_-<k_0+\omega
  <\omega_+)  \nn&& +\,(n(\omega-k_0) -n(\omega))\Theta_\Gamma(\omega_-<
  \omega-k_0<\omega_+) \bigr].\nn
\label{imsab}
\end{eqnarray}
This shows the other expected effect conjectured in the Introduction,
the enlarged phase space contributing to the damping, since
$\Theta_\Gamma$ is a smeared version of $\Theta$. Next we present some
direct consequences of the application of eq.(\ref{imsab}).

\subsection{On-shell damping rate of Higgs bosons}

If the coefficient of $g^2T$ in the perturbative expression of the
damping rate is ${\cal O}(1)$, then the corrections from the finite
width are negligible, or more precisely they cannot be assessed
reliably from the present approximation. In these cases the Higgs mass
is the relevant scale which determines the damping rates, the finite
width plays only subleading role. As an example let us take the
on-shell damping of the Higgs component. This phenomenon is the result
of the decay of the Higgs boson into two Goldstone bosons in the zero
width case. Now we want take into account the effect of broadening of
the resonances in this process. Using (\ref{gammagap}), (\ref{impirs})
and the first line of (\ref{imsab}) we write ($k_0^2=\k^2+m_H^2$)
\begin{eqnarray}
  \gamma_{1,on-shell}(\k) = \frac{g^2 m_H^2(N-1)}{198\pi k_0|\k|}
  \int\limits_0^\infty\!dp\!&& (1+n(p)+n(k_0-p))\nn&&
  \times\Theta_\Gamma(|p-|\k|| < k_0-p < p+|\k|). 
\end{eqnarray}
If $|\k|\to 0$ (and $p>|\k|$) then using eq.~(\ref{comptheta})
\begin{equation}
  \lim_{|\k|\to 0} \frac1{|\k|} \Theta_\Gamma(\omega_- < s < \omega_+)
   = 2\frac{d\omega_b(p)}{dp}
   \frac{d\Theta_\Gamma(\omega_b-s)}{d\omega_b} = 
   \frac{2p}{\omega_b} \delta_\Gamma(\omega_b-s).
\label{smallk}
\end{equation}
Now $\omega_b=p$ and $s=m-p$, therefore
\begin{equation}
   \gamma_{1,on-shell}(\k\to0) = \frac{g^2 m_H(N-1)}{96\pi}
   (1+2 n(\frac{m_H}2)) \int\limits_0^\infty\!dp\,
   \delta_\Gamma(m_H-2p).
\label{gamma1}
\end{equation}
The integral is formally $(1-\Gamma(m_H)/m_H)/2$, however, within our
approximation scheme it can be also be approximated by $1/2$ (by the
rule of the application of $\delta_\Gamma$ to a smooth function). The
difference can be tracked back to the neighborhood of the $p=0$ lower
limit of the integration, which yields $\delta_\Gamma(m_H)$. After
closer inspection of the two-particle cut contributions to this
process, it can be seen that the far off shell contributions (for
example the ones coming from $S_{11}$) are of the same order of
magnitude. As a consequence we cannot give the correction to the
one-loop result for $\gamma_{1,on-shell}(\k)$, as it was stated above.
The conclusion is that better approximation scheme is needed to the
solution of the self-consistent equations in order to improve it
beyond the one-loop result in case of Higgs bosons.

\subsection{Landau damping and on shell damping rate of Goldstone bosons}

The situation is different in case of the on-shell damping of
Goldstone particles at small momenta. From perturbation theory
\cite{ourpaper}
\[ \gamma_{i,on-shell}^{pert}(\k\to0) \sim \frac1{|\k|} \exp\left(-
  \frac{m_H^2}{ 4 |\k| T}\right),\] 
which is exponentially small. Here it makes sense to examine, how its
self-energy is modified due to the quasi-particle width. In our
approximation, using (\ref{gammagap}), (\ref{impirs}) and the Landau
damping part of (\ref{imsab})
\begin{eqnarray}
   \Im\Pi_i(k) = - \frac{g^2 m_H^2}{48\pi |\k|}\int
   \limits_0^\infty \! dp \, \biggl[ && (n(p)-n(p+k_0))
   \Theta_\Gamma(\omega_- < p+k_0 < \omega_+) \nn&& + (n(p-k_0)-n(p))
   \Theta_\Gamma(\omega_- < p-k_0 < \omega_+)\biggr].
\end{eqnarray}
If $|\k|\ll m_h,T$ we can use eq.~(\ref{smallk}) and write
($\omega^2=\k^2+m_H^2$)
\begin{eqnarray}
  \Im\Pi_i(k) = - \frac{g^2 m_H^2}{24\pi}\int
  \limits_0^\infty \! \frac{dp\,p}{\omega} \, \biggl[
  &&(n(p)-n(p+k_0)) \delta_\Gamma(\omega- p- k_0)\nn&& +
  (n(p-k_0)-n(p)) \delta_\Gamma(\omega- p +k_0)\biggr].
\end{eqnarray}
If also $k_0\ll m_H,T$ is fulfilled we expand the Bose-Einstein
factors in powers of their arguments, and find for the leading term
\begin{equation}
  \Im\Pi_i(k) = - \frac{g^2 m_H^2 k_0}{12\pi}\int\limits_0^\infty \!
  \frac{dp\,p}{\omega} \, (-\frac{dn}{dp})\delta_\Gamma(\omega- p).
\end{equation}
For $\Gamma=0$ the delta function can never be satisfied. For large
$p$, however, $\omega-p\approx m_H^2/(2p)$, the argument of
$\delta_\Gamma$ approaches zero (unlike the previous cases, where the
argument was constant).  Moreover, for large $p$, $\Gamma\sim
-\Im\Pi/2p$, remains smaller than $\omega-p$. Assuming that this
region dominates the integral (to be confirmed a posteriori) we
neglect $\Gamma$ in the denominator and write
\begin{equation}
  \Im\Pi_i(k) = - \frac{g^2 k_0}{3\pi^2 m_H^2} \int\limits_0^\infty \!
  dp\,\Gamma\, \frac{p^3}{\omega} (-\frac{dn}{dp}).
\end{equation}
The integrand receives contribution from momenta $p\sim T>m_H$,
therefore the starting assumption was correct. Approximating
$\Gamma$ by a constant, the integral can be performed
\begin{equation}
  \Im\Pi_i(k) = - \frac{g^2T^2 k_0 \Gamma}{ 9 m_H^2}  = -\eta_i\, k_0.
\end{equation}
Here $\Gamma = \gamma_1(p) + \gamma_i(p)$ must be taken at the
dominating value of the momentum $p \sim T$. Our notation emphasizes
that a viscosity term has appeared, similar to the one observed in
gauge theories \cite{gaugevisc}. According to our result the viscosity
coefficient $\eta$ is proportional to $\Gamma(T)$, the quasi-particle
damping about $T$. The above formula also says that long wavelength
Goldstone modes are damped in contrary to the zero width result, and
\begin{equation}
  \gamma_{i,on-shell}(\k=0) = \frac{\eta_i}2.
\end{equation}
If we take for $\Gamma$ only the Higgs contribution, as calculated
from perturbation theory \cite{ourpaper}
\begin{equation}
  \eta_i \approx \frac{g^4 T (N-1)}{432 \pi} \ln\frac T{m_H}.
\end{equation}

\subsection{Higgs Landau damping}

Landau damping of the Higgs fluctuation modes in the usual
perturbation theory comes from the $k^2<0$ contributions of $S_{11}$
and $S_{ii}$. We now examine their modification due to the finite
lifetime of quasi-particles.

First let us consider $S_{11}$. We will apply high temperature
expansion on the results. More precisely in the last two lines in
eq.~(\ref{imsab}) we assume $p\gg m,|\k|$, thus neglect masses and
$\omega_\pm \approx p\pm|\k|$. Thus
\begin{equation}
  \Theta_\Gamma(\omega_- < \omega \pm k_0 < \omega_+) \approx
  \Theta_\Gamma(k_0+ |\k|) -  \Theta_\Gamma(k_0 - |\k|) =
  \Theta_\Gamma(-|\k|<k_0<|\k|),
\end{equation}
depends on $\omega$ only through $\Gamma$. If $\Gamma=0$ this is
nonzero only for $k_0<|\k|$ and the usual abrupt end of the Landau
damping phenomenon occurs for $k_0=|\k|$. For finite width, however,
the range of effect in $k_0$ might broaden. The Landau damping part of
$\Im S_{11}$ then reads
\begin{equation}
  \Im S_{11}^{Land.damp.}\approx -\frac1{16\pi|\k|}
  \int\limits_{m_H}^\infty \!d\omega\, (n(\omega-k_0) - n(\omega+k_0))
  \Theta_\Gamma(-|\k|<k_0<|\k|).
\end{equation}
The smooth difference $n(\omega-k_0) - n(\omega+k_0) \approx -2k_0
dn/d\omega$ enhances contributions from small momenta. Therefore if we
approximate $\Gamma(p) = 2\gamma_1(p)$ by a constant, we have to take
it at a typical momentum $p\approx 0$. From perturbation theory (see
eq.~(\ref{gamma1}))
\begin{equation}
  \Gamma\approx 2\gamma_1(0) \approx \frac{g^2 T}{24\pi}.
\end{equation}

Then high temperature expansion yields
\begin{eqnarray}
  \Im S_{11}^{Land.damp.}&&\approx -\frac T{16\pi|\k|} \ln\frac{m_H+k_0}
  {m_H-k_0} \Theta_\Gamma(-|\k|<k_0<|\k|)
  \nn&& \approx -\frac {T}{8\pi m_H} \frac{k_0}{|\k|}
  \Theta_\Gamma(-|\k|<k_0<|\k|).
\label{s11landam}
\end{eqnarray}
This is the zero width result simply modified by the substitution
$\Theta(|\k| -k_0) \to   \Theta_\Gamma(-|\k|<k_0<|\k|)$.

In case of the Landau damping contribution from $\Im S_{ii}$ the
situation is complicated by IR divergences. To be on the safe side we
introduce there an IR cutoff $|\k|< \Lambda < m_H$, and use cut
Bose-Einstein distributions \cite{ourpaper}
\[\tilde n(\omega) = \Theta(\omega-\Lambda) n(\omega).\]
The result of the calculation is thereafter interpreted as induced
self-energies for an effective model for modes below the cutoff
\cite{gaugevisc,ourpaper}. Then the Goldstone modes do not
contribute, and the Landau damping part of the Higgs self-energy reads
(cf. (\ref{impirs}))
\begin{equation}
  \Im\Pi_1^{R,Land,damp}(k) = -\frac{3g^2}{16\pi}\,
  T\,m_H\,\frac{k_0}{|\k|}   \Theta_\Gamma(-|\k|<k_0<|\k|).
\end{equation}
Qualitatively new phenomenon appears in comparison to the zero width
case if $k_0,|\k| < \Gamma$, ie. below the scale $\sim g^2
T$. There (cf. (\ref{smtheta})) $\Theta_\Gamma(-|\k|<k_0<|\k|) \approx
2|\k|/(\pi\Gamma)$, and
\begin{equation}
  \Im\Pi_1^{R,Land,damp}(|\k|,k_0<\Gamma) = -\frac{3g^2}{8\pi^2}\,
  \frac{T m_H}{\Gamma}\,k_0 = -\eta_1 k_0.
\end{equation}
This is independent of \k\ (also the usual kinematical restriction for
the Landau damping, $k_0<|\k|$ does not hold), and is valid till
$k_0<\Gamma$. If we consider it as part of the effective action for
the low frequency modes, in real space this turns into an $\sim i\d_t$
viscosity term.  The presence of a similar effect has been demonstrated
in gauge theories \cite{gaugevisc}.  The present calculation shows that
this phenomenon is common in effective dynamical theories of the low
frequency modes of any non-linear field theory.

\section{Conclusions}

In this paper we made an attempt to incorporate higher loop effects
into the perturbation theory and work out some of the most dramatic
consequences they might provide. We expect, on general grounds, two
types of effects. The first is the result of the larger available
phase space for the processes contributing to the damping which
increase the physical damping rates. The other effect is due to the
instability of intermediate particles which lead to loss-of memory
effects.

We have demonstrated, how to resum self-consistently a subset of
Feynman diagrams (self-energy corrections), what leads to the closed
set of functional equations (gap equations) of eq.~(\ref{gapeq}). Its
general solution is very complicated, therefore we have made some
simplifying assumptions. Among them the most important was the
Breit-Wigner quasi-particle approximation. Then the formalism is
similar to the zero-width calculations, just the Dirac-delta and the
Theta distributions are replaced by smeared versions. We have
calculated the imaginary part of the self-energies in this
approximation.

We have shown, how the Landau damping regimes of the different
excitations are modified due to the finite width effects. In general
we have arrived at a formula for $k_0,|\k|\ll \gamma$ 
\begin{equation}
  \Im\Pi_a^{\textrm{\small Landau\ damping}}(k) = -\eta_a k_0,
\end{equation}
which corresponds to $\eta_a \d_t$ in the real space. Such a viscosity
term appeared also in the gauge theories \cite{gaugevisc}. This again
confirms our expectation about the loss-of-memory effects beyond the
time scale $\sim1/\gamma$. Long wavelength Goldstone bosons
($k_0=|\k|\ll\gamma$) fall into this regime, thus having a finite
damping rate $\gamma_i^{on-shell}(\k) =\eta_i/2$, where $\eta_i\sim
\gamma_1(T)$, the on-shell damping rate of the Higgs bosons with
momentum about $T$. Other on-shell damping rates which in the
perturbation theory are proportional to $g^2T$ with a coefficient of
${\cal O}(1)$ receive ${\cal O}(\gamma/m)\sim g$ correction, which is,
however, beyond the scope of the present approximation.

After this calculation one faces the question still awaiting
clarification, which subset of Feynman diagrams yields the correct
time evolution of the physical quantities. We know, for example, that
the $k_0=0$ Goldstone mode which corresponds to the homogeneous
rotation of the vacuum is not damped (cf. \cite{ourpaper,BoPaSz}) in
contradiction with the present result.  Probably this puzzle can be
resolved by a resummation of a subset of vertex diagrams, similarly to
the resummation in the $\Phi^4$ theory \cite{ladderresum} which leads
to an important modification in the viscosity coefficients. It is also
worth to study, how the present ideas can be applied to gauge theories
and to systems far from equilibrium.

\begin{appendix}

\section*{Appendix}
\label{sec:appa}

The appearance of retarded self-energy can be determined from the
Schwinger-Dyson equations for the matrix propagator. We will use here
the matrix notations, $G^{11}=G^c,\, G^{22}=G^a,\, G^{12}=G^<$ and
$G^{21}=G^>$
\begin{equation}
  G^{ij}=G^{0,ij} + G^{0,ik}\Pi^{kl}\,G^{lm},
\end{equation}
where the indices can be 1 or 2. Writing out the components of these
equations one finds
\begin{eqnarray}
  && G^{i1}= G^{0,i1} + G^{0,i1}\Pi^{1j}\,G^{j1} +
  G^{0,i2}\Pi^{2j}\,G^{j1} \nn
  && G^{i2}= G^{0,i2} + G^{0,i1}\Pi^{1j}\,G^{j2} +
  G^{0,i2}\Pi^{2j}\,G^{j2}.
\end{eqnarray}
Subtracting the two equations and using $G^R=G^{j1}-G^{j2}$ for both
$j=1,2$ we find
\begin{equation}
  G^R=G^{0,R} + G^{0,i1}(\Pi^{11}+\Pi^{12})\,G^R +
  G^{0,i2}(\Pi^{21}+\Pi^{22})\,G^R.
\end{equation}
Subtracting again the equaitons $i=1$ and $i=2$, then multiplying the result
by $(G_R)^{-1}$ and using $G^{11}-G^{21}=G^{12}-G^{22}$ follows
\begin{equation}
  \Pi^{11}+\Pi^{12}+\Pi^{21}+\Pi^{22}=0.
\end{equation}
Introducing $\Pi^R=\Pi^{11}+\Pi^{12}$ we arrive finally at
\begin{equation}
  G^R=G^{0,R} + G^{0,R}\Pi^R\,G^R,
\end{equation}
or
\begin{equation}
  (G^R)^{-1} = (G^{0,R})^{-1} -\Pi^R
\end{equation}
Schwinger-Dyson equations.

The formula under (\ref{intformula}) can be derived in the following
way. For $\gamma>0$
\begin{eqnarray}
  && \frac1{2\pi}\int\limits_{-\infty}^\infty\frac{dx}{(x+i\alpha)
    (\Omega -x +i\gamma)} \nn
  &&=\frac1{2\pi}\, \frac1{\Omega+ i(\alpha+\gamma)}
  \int\limits_{-\infty}^\infty\!dx \left[ \frac1{x+i\alpha} -
    \frac1{x-\Omega-i\gamma}\right] = \nn
  &&= \frac1{2\pi}\, \frac1{\Omega+ i(\alpha+\gamma)} \left[
    \int\limits_{-\infty+i\alpha}^{\infty+i\alpha}\!dx -
    \int\limits_{-\infty-i\gamma}^{\infty-i\gamma}\!dx\right]
  \frac1x\nn
  && = \frac1{2\pi}\, \frac1{\Omega+ i(\alpha+\gamma)} \left[
    \ln(-1-i\alpha 0^+) -\ln(-1+i\gamma 0^+)\right] =\nn
  &&=  \Theta(\alpha)\,\frac{-i}{\Omega+ i(\alpha+\gamma)}.
\end{eqnarray}
On the other hand
\begin{eqnarray}
  \frac1{2\pi}\int\limits_{-\infty}^\infty\frac{dx}{(x+i\alpha)
    (\Omega -x - i\gamma)} &&= \frac1{2\pi}\int\limits_{-\infty}^\infty
    \frac{dx}{(x - i\alpha)(- \Omega -x +i\gamma)} \nn&&=
    \Theta(-\alpha)\,\frac{i}{\Omega+ i(\alpha-\gamma)}.
\end{eqnarray}
The two equations lead to the result of eq~(\ref{intformula}).

\end{appendix}

\section*{Acknowledgements}
I am grateful to Z. Fodor, P.Petreczky, Zs. Sz\'ep and especially to
A. Patk\'os for discussions and comments on the manuscript.  This work
was partially supported by OTKA (Hungarian Science Fund).

\end{document}